\RequirePackage{fix-cm}
\documentclass[smallextended]{svjour3}
\smartqed
\usepackage{graphicx}

\let\textcite\citet

\makeatletter
\let\cl@chapter\undefined
\makeatletter

\usepackage[T1]{fontenc}
\usepackage{url}
\usepackage{amsmath}
\usepackage[table,xcdraw,dvipsnames]{xcolor}
\usepackage{colortbl}
\usepackage{enumitem}
\usepackage{xspace}
\usepackage{soul}
\usepackage{hhline}
\usepackage{natbib}

\usepackage{hyperref}
\hypersetup{hidelinks}
\usepackage[nameinlink,capitalize]{cleveref}

\usepackage[caption=false]{subfig}
\usepackage{multirow}
\usepackage{adjustbox}
\usepackage{rotating}
\graphicspath{{figures/}}
\usepackage{graphicx}
\usepackage{booktabs}
\usepackage[utf8]{inputenc}

\usepackage{listings}
\usepackage{mathtools}

\makeatletter
\lst@Key{countblanklines}{true}[t]%
    {\lstKV@SetIf{#1}\lst@ifcountblanklines}

\lst@AddToHook{OnEmptyLine}{%
    \lst@ifnumberblanklines\else%
       \lst@ifcountblanklines\else%
         \advance\c@lstnumber-\@ne\relax%
       \fi%
    \fi}
\makeatother

\definecolor{lightgreen}{cmyk}{0.21,.03,.29,0}
\definecolor{codeblack}{rgb}{0,0,0}
\definecolor{lightgray}{gray}{0.98}
\definecolor{codeblue}{rgb}{0.13,0.13,1}
\definecolor{codegrey}{rgb}{0.36,0.35,0.38}
\definecolor{codegreen}{rgb}{0,0.5,0}
\definecolor{codered}{rgb}{0.9,0,0}
\definecolor{OliveGreen}{cmyk}{0.64,0,0.95,0.40}
\definecolor{DarkGreen}{cmyk}{0.40,0,0.50,0.15}
\definecolor{purple}{cmyk}{0.41,0.73,0,0.1}
\definecolor{LightRed}{cmyk}{0,0.682,0.728,0.6}

\lstdefinestyle{default}{
    escapechar=\$,
	backgroundcolor=\color{lightgray},
	keywordstyle=[1]\color{codeblue},
	keywordstyle=[2]\color{purple},
	keywordstyle=[3]\color{codered},
 	stringstyle=\color{codegreen},
	commentstyle=\color{codegrey},
	basicstyle=\color{codegrey}\footnotesize\ttfamily,
	columns=fullflexible,
    breaklines=true,
	prebreak=\raisebox{0ex}[0ex][0ex]{\ensuremath{\hookleftarrow}},
    xleftmargin=6pt,
    xrightmargin=0pt, 
	frame=none,
    framexleftmargin=0pt,
    framexrightmargin=0pt,
    mathescape=true,
	numbers=left,
	numberblanklines=false,
    countblanklines=false,
	stepnumber=1,
	numbersep=4pt,
	numberstyle=\footnotesize,
	keepspaces=true,
    showspaces=false,
    showstringspaces=false,
	showtabs=false,
	upquote=true,
    tabsize=2
}

\lstnewenvironment{Java}
  {\lstset{language=Java,style=default}}
  {}

\newcommand{\highlight}[1]{\cellcolor{lightgreen!50}\textbf{#1}\xspace}

\usepackage{enumitem}
\newlist{rqs}{enumerate}{1}
\setlist[rqs]{label={\textbf{RQ\arabic*}},align=parleft,left=0em..3em,ref={RQ\arabic*},resume}
\newlist{enumline}{enumerate*}{1}
  \setlist[enumline]{label=(\roman*),itemjoin={{, }},itemjoin*={{; and }}}

\usepackage{mdframed}
\usepackage{xpatch}

\makeatletter
\xpatchcmd{\endmdframed}
  {\aftergroup\endmdf@trivlist\color@endgroup}
  {\endmdf@trivlist\color@endgroup\@doendpe}
  {}{}
\makeatother
\newcounter{FindingCounter}
\stepcounter{FindingCounter}
\newcommand{\findings}[1]{
	\begin{mdframed}[backgroundcolor=gray!10,
			linewidth=0.75pt,
			roundcorner=5pt,
			innertopmargin=2mm,
			innerbottommargin=2mm,
			innerrightmargin=2mm,
			innerleftmargin=2mm,
			skipabove=1mm,
			skipbelow=1mm,
			font=\small]
		{\bf{Finding~\arabic{FindingCounter}:}}~ #1
	\end{mdframed}
	\stepcounter{FindingCounter}
}

\usepackage{lmodern}
\graphicspath{{figures}}
\definecolor{seaborncolorblind1}{RGB}{1,115,178}
\definecolor{seaborncolorblind2}{RGB}{222,143,5}
\definecolor{seaborncolorblind3}{RGB}{2,158,115}
\definecolor{seaborncolorblind4}{RGB}{213,94,0}
\definecolor{seaborncolorblind5}{RGB}{204,120,188}
\definecolor{seaborncolorblind6}{RGB}{202,145,97}
\definecolor{seaborncolorblind7}{RGB}{251,175,228}
\definecolor{seaborncolorblind8}{RGB}{148,148,148}
\definecolor{seaborncolorblind9}{RGB}{236,225,51}
\definecolor{seaborncolorblind10}{RGB}{86,180,233}

\usepackage{tikz}
\usetikzlibrary{shapes,arrows.meta,positioning,quotes}

\usepackage{comment}
\usepackage{pdflscape}

\newcounter{customlstnum}
\newcommand{\pfaszz}{WIA-SZZ\xspace}

\newcommand{\ocfilter}{One Commit Filter\xspace}
\newcommand{\issuefilter}{Issue Filter\xspace}

\begin{document}

\title{\pfaszz: Work Item Aware SZZ}
\titlerunning{\pfaszz}

\author{Salom\'{e} Perez-Rosero         \and
Robert Dyer                             \and
Samuel W. Flint                         \and
Shane McIntosh                          \and
Witawas Srisa-an
}

\institute{S. Perez-Rosero, R. Dyer, S. W. Flint, W. Srisa-an \at
              School of Computing \\
              University of Nebraska--Lincoln, USA \\
              \email{\{mperezrosero2,rdyer,swflint,witawas\}@huskers.unl.edu}
           \and
           S. McIntosh \at
              David R. Cheriton School of Computer Science\\
              University of Waterloo, Canada \\
              \email{shane.mcintosh@uwaterloo.ca}
}

\date{Received: date / Accepted: date}

\maketitle

\abstract{Many software engineering maintenance tasks require linking a commit that induced a bug with the commit that later fixed that bug. Several existing SZZ algorithms provide a way to identify the potential commit that induced a bug when given a fixing commit as input.  Prior work introduced the notion of a ``work item'', a logical grouping of commits that could be a single unit of work.  Our key insight in this work is to recognize that a bug-inducing commit and the fix(es) for that bug together represent a ``work item.''  It is not currently understood how these work items, which are logical groups of revisions addressing a single issue or feature, could impact the performance of algorithms such as SZZ.  In this paper, we propose a heuristic that, given an input commit, uses information about changed methods to identify related commits that form a work item with the input commit.  We hypothesize that given such a work item identifying heuristic, we can identify bug-inducing commits more accurately than existing SZZ approaches.  We then build a new variant of SZZ that we call Work Item Aware SZZ (WIA-SZZ), that leverages our work item detecting heuristic to first suggest bug-inducing commits.  If our heuristic fails to find any candidates, we then fall back to baseline variants of SZZ.  We conduct a manual evaluation to assess the accuracy of our heuristic to identify work items. Our evaluation reveals the heuristic is 64\% accurate in finding work items, but most importantly it is able to find many bug-inducing commits.  We then evaluate our approach on 821 repositories that have been previously used to study the performance of SZZ, comparing our work against six SZZ variants.  That evaluation shows an improvement in $F_1$ scores ranging from 2\% to 9\%, or when looking only at the subset of cases that found work item improved 3\% to 14\%.}

\maketitle

\section{Introduction}
\label{sec:intro}

The ability to accurately identify which commit(s) introduced a bug into a system provides many benefits to researchers.  Knowing about past bug-inducing commits (BICs) allows researchers to potentially predict if new commits are introducing similar bugs~\citep{aversano07,fukushima14,jit,kim08,kim07,kim13,yang15}, use BICs during debugging to make it easier to debug~\citep{an2023fonte,WenFSE19}, understand what factors cause the introduction of bugs~\citep{Asaduzzaman12,Bernardi12,Bavota12}, etc.

One of the leading approaches to automatically identify BICs is the SZZ algorithm~\citep{szz}. SZZ works by taking a commit identified as a fixing commit (FC), examining the diff of that commit, and implicating past commits that modified similar lines as potentially being the BIC. This assumes the FC given as input is a good starting point, which may not always be the case since fixes can often span multiple commits. \cite{park12} found that 22-32\% of resolved bug reports required a supplemental fix, while \cite{an14} found that 10-26\% required a supplemental fix.

SZZ approaches often use information about the reported bug and filter candidate BICs that appear between the issue date and the FC. \cite{rezk22} recently showed that can tend to remove valid candidates, as 0.35--14.49\% (median 5.46\%) of the filtered candidates are actually defect-fixing commits themselves.  Around a third of those removed commits are due to an overly aggressive issue date filter, which often affects fixes with supplemental fixes linked to the same issue ID.

An investigation by \cite{WenFSE19} found a strong relationship between BIC and FC: 73.2\% of bug-inducing code and bug repairs occur in the same source files. They also found that 64.6\% of repair statements can be traced from the statements that induced bugs. Thus, it may be possible to identify related commits that attempt to fix bugs by tracing activities within each commit from BIC to FC and vice versa.

To help identify these related commits, \cite{mcintosh11:_work_items} introduced the notion of ``work items'', defined as logical groupings of commits that add a feature or fix a bug.  By that definition, supplemental and reopened bugs could also be classified as work items.  \cite{miura16:_work_items} found in half the systems they studied, 29\% of the work items consisted of at least two commits. While there are many SZZ variants in the literature~\citep{szz,ra-szz,ag-szz,l-szz,ma-szz,rastar-szz,bludau2022pr} and other approaches for identifying bug-inducing commits~\citep{an2023fonte}, none have investigated the impact work items could have on the accuracy of their approach and instead, they assume the commit given as the fix contains the full bugfix.

\begin{figure*}[ht]
  \centering
  \includegraphics[width=\linewidth]{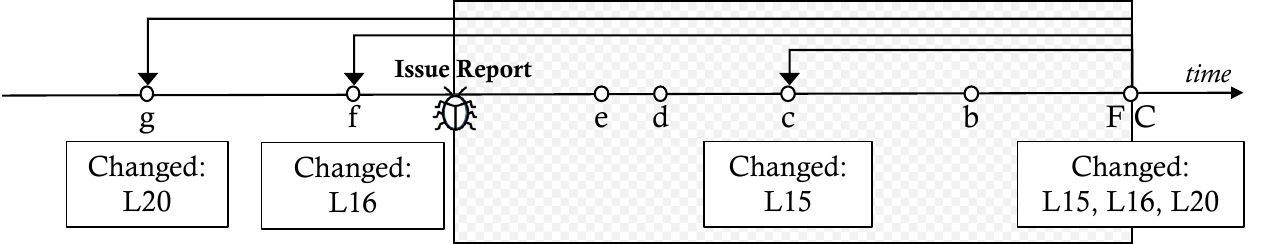}
  \caption{Overview of SZZ~\citep{szz}, where the lines changed at the fix commit (FC) are annotated with a tool such as Git blame to show the previous commits that modified those lines.  In this example, FC modified three lines and those lines were most recently modified in commits \texttt{c}, \texttt{f}, and \texttt{g}.  If an issue report is provided, the date of that report can create a filter that removes anything in the checkered region (in this case, commit \texttt{c}). Thus, FC implicates commits \texttt{f} and \texttt{g} as possibly inducing the bug.}
  \label{fig:szz-overview}
\end{figure*}

In this work, we present a heuristic to detect potential work items by leveraging the relationship between FC and BIC, tracking common changes at the method-level granularity. We chose to model the FC-BIC relationship by tracking changes at the method level, inspired by the work of \cite{pan_09}. In that study, the authors examined fix-bug patterns from 7 open source Java projects and classified them into 9 different categories: if-related, method call, loop, assignment, switch, try, method declaration, sequence, and class field. Code modifications across these categories occur at the inner method body; thus, we selected methods as the granularity to build our heuristic. Once our heuristic leverages work items from the FC, we use its output with standard SZZ algorithms to detect bug-inducing commits. Our approach, 
\emph{Work Item Aware SZZ} (\pfaszz), first attempts to identify commits that might be work items for a given fix commit.
If no work item commits are found, or all such commits are newer than the issue report date, then we fall back to pre-existing SZZ algorithms. 

We evaluate \pfaszz on a previously used dataset~\citep{rosa21}, showing first that our algorithm identifies 54\% of all fix commits as being part of a work item.  For those commits, our approach significantly outperforms the six studied SZZ variants~\citep{szz,ra-szz,ag-szz,l-szz,ma-szz} with a 3--14\% improvement in $F_1$-score.  

Since \pfaszz uses \ocfilter to identify a single BIC, unlike other SZZ variants which identify multiple BICs, we also study the extent to which this filter may cause over-fitting.
To do this, we evaluate the performance of other variants with and without this filter, observing that the \ocfilter does not improve variants, indicating it does not cause overfitting.
Finally, we investigate the effect of applying a filter based on the issue date and show that our approach performs substantially better with that filter in place.

This paper makes the following contributions:

\begin{itemize}
    \item Provides a heuristic-based algorithm to identify work items in historical commit records.
    
    \item First application of the notion of work items to identify bug-inducing commits, showing an improvement of 3--14\% over existing SZZ variants.

    \item First empirical evidence that shows bugs and their fixes are work items.
\end{itemize}

In the next section, we briefly overview the SZZ algorithm.  In \autoref{sec:related} we discuss closely related works.  Then in \autoref{sec:approach} we describe a heuristic for detecting work items and our custom SZZ algorithm.  We evaluate our approach in \autoref{sec:eval}.  We then conclude in \autoref{sec:conclusion}.

\section{Background}
\label{sec:bg}

In this section, we provide some background information about the SZZ algorithm, the definition of work items, and their relation to bug-inducing commits.

\subsection{Overview of SZZ}

First, we briefly describe the original SZZ algorithm~\citep{szz}.  Note that many other variants exist, which we briefly describe in the related works, but they all operate similarly to the original algorithm.

SZZ requires a fixing commit (FC) that addresses a previous bug to identify the bug-inducing commit (BIC). It analyzes the lines of code modified or removed in the FC, using VCS annotation tools like Git's blame/annotate, to trace these changes back in time. This yields a set of candidate commits potentially introducing the bug. Optionally, specifying the related issue report allows filtering out commits made after the issue report's date, as the bug was already known then. For instance, given an FC with changes on lines 15, 16, and 20 (as shown in \Cref{fig:szz-overview}), a Git blame implicates commits \texttt{c}, \texttt{f}, and \texttt{g}. Without additional context, SZZ would consider all three commits as potentially inducing the bug; however, with the issue report, commit c could be filtered out as it postdates the bug report.

\subsection{Work Items}

\begin{table*}[ht]
\centering
\definecolor{diffbg}{gray}{0.96}
\definecolor{diffaddbg}{rgb}{0.7333333333,0.8745098039,1}
\definecolor{diffadd}{rgb}{0.9333333333,0.9803921569,1}
\definecolor{diffremovebg}{rgb}{0.9882352941,0.8392156863,0.6901960784}
\definecolor{diffremove}{rgb}{1,0.9803921569,0.9568627451}
\newcommand{\lineno}[1]{\cellcolor{diffbg}\color{gray}#1}
\newcommand{\ind}[0]{\hspace{0.5em}}

\caption{
Example from a real-world repository (\href{https://archive.softwareheritage.org/browse/origin/directory/?origin_url=https://github.com/nickvandewiele/RMG-Java&timestamp=2021-01-18T04:37:46Z}{nickvandewiele/RMG-Java}) from \cite{rosa21} with code snippets for FC (\texttt{\href{https://archive.softwareheritage.org/browse/revision/8c5f991e6876de001ff11829ceb9894d11c80014/?origin_url=https://github.com/nickvandewiele/RMG-Java&snapshot=123dd32e917eb3ab0dcc84f7fb83a06532a0c37d&timestamp=2021-01-18T04:37:46Z\#swh-revision-changes}{8c5f991e6}}) and one work item (\texttt{\href{https://archive.softwareheritage.org/browse/revision/7339cc176cc7bbc671f4dec65b2f161d351ea270/?origin_url=https://github.com/nickvandewiele/RMG-Java&snapshot=123dd32e917eb3ab0dcc84f7fb83a06532a0c37d&timestamp=2021-01-18T04:37:46Z\#swh-revision-changes}{7339cc176}}). This is a case of code modifications with only insertions at the FC, where the FC adds three lines (of special interest line~887) to complement the logic of the else branch initially inserted in the BIC. The work item finding heuristic described in this paper finds the BIC from the FC without false positives. None of the other SZZ variants work in the case of a commit with only inserted code.}

\renewcommand{\arraystretch}{1.3}
\renewcommand{\tabcolsep}{0.9pt}
\resizebox{\textwidth}{!}{%
\begin{tabular}{|rcl|rcl|}
\hline
\multicolumn{6}{|c|}{\bf source/RMG/jing/chem/Species.java } \\
\multicolumn{6}{|c|}{\bf public static Species make (String p\_name, ChemGraph p\_chemGraph) } \\
\hline
\multicolumn{3}{|c}{\bf\scriptsize FC (8c5f991e6)} & \multicolumn{3}{c|}{\bf\scriptsize WI -- BIC (7339cc176)} \\
\hline

\lineno{877} & \lineno{~} & \cellcolor{diffbg}{if ((p\_name == null || p\_name.length()==0) \&\& ...)\{} & \lineno{\cellcolor{diffaddbg}880} & \cellcolor{diffaddbg}+ & \cellcolor{diffadd}\ind \colorbox{diffaddbg}{if ((p\_name == null || p\_name.length()==0) \&\& ...)\{} \\

\lineno{878} & \lineno{~} & \cellcolor{diffbg}{\ind name = spe.getThermoData().getName();} & \lineno{\cellcolor{diffaddbg}881} & \cellcolor{diffaddbg}+ & \cellcolor{diffadd}\ind \colorbox{diffaddbg}{\ind name = spe.getThermoData().getName();} \\

\lineno{879} & \lineno{~} & \cellcolor{diffbg}{\ind if (name.matches("s//d\{8\}")) \{ // ...} & \lineno{\cellcolor{diffaddbg}882} & \cellcolor{diffaddbg}+ & \cellcolor{diffadd}\ind \colorbox{diffaddbg}{\ind if (name.matches("s//d\{8\}")) \{ // ...} \\

\lineno{880} & \lineno{~} & \cellcolor{diffbg}\ind \ind // ... & \lineno{\cellcolor{diffaddbg}883} & \cellcolor{diffaddbg}+ & \cellcolor{diffadd}\ind \colorbox{diffaddbg}{\ind \ind // ...} \\

\lineno{881} & \lineno{~} & \cellcolor{diffbg}\ind \} else \{ & \lineno{\cellcolor{diffaddbg}884} & \cellcolor{diffaddbg}+ & \cellcolor{diffadd}\ind \colorbox{diffaddbg}{\ind \} else \{} \\

\lineno{882} & \lineno{~} & \cellcolor{diffbg}\ind \ind // ... & \lineno{\cellcolor{diffaddbg}885} & \cellcolor{diffaddbg}+ & \cellcolor{diffadd}\ind \colorbox{diffaddbg}{\ind \ind // ...}\\

\lineno{883} & \lineno{~} & \cellcolor{diffbg}\ind \ind spe.setName(spe.getThermoData().getName()) & \lineno{\cellcolor{diffaddbg}886} & \cellcolor{diffaddbg}+ & \cellcolor{diffadd}\ind \colorbox{diffaddbg}{\ind \ind spe.setName(spe.getThermoData().getName());} \\

\lineno{\cellcolor{diffaddbg}884} & \cellcolor{diffaddbg}+ & \cellcolor{diffadd}\ind \colorbox{diffaddbg}{// ...} & \cellcolor{diffbg}~ & \cellcolor{diffbg}~ & \cellcolor{diffbg}\ind ~ \\

\lineno{\cellcolor{diffaddbg}885} & \cellcolor{diffaddbg}+ & \cellcolor{diffadd}\ind \colorbox{diffaddbg}{// ...} & \cellcolor{diffbg}~ & \cellcolor{diffbg}~ & \cellcolor{diffbg}\ind ~ \\

\lineno{\cellcolor{diffaddbg}886} & \cellcolor{diffaddbg}+ & \cellcolor{diffadd}\ind \colorbox{diffaddbg}{spe.generateNASAThermoData();} & \cellcolor{diffbg}~ & \cellcolor{diffbg}~ & \cellcolor{diffbg}\ind ~ \\

\lineno{887} & \cellcolor{diffbg}~ & \cellcolor{diffbg}\ind \colorbox{diffbg}{\ind \}} & \lineno{\cellcolor{diffaddbg}887} & \cellcolor{diffaddbg}+ & \cellcolor{diffadd}\ind \colorbox{diffaddbg}{\ind \}} \\

\lineno{888} & \lineno{~} & \cellcolor{diffbg}\ind \} & \lineno{\cellcolor{diffaddbg}888} & \cellcolor{diffaddbg}+ & \cellcolor{diffadd}\ind \colorbox{diffaddbg}{\}}\\

\hline
\end{tabular}
}
\label{tab:code-example}
\end{table*}

\cite{mcintosh11:_work_items} introduced the notion of ``work items'', a logical grouping of commits contributing to the same feature or bug fix. The notion is useful for SZZ algorithms, as typical SZZ algorithms require an FC as the starting point and attempt to trace back toward a BIC.  Since the FC and the BIC are logically working toward the same feature~\citep{WenFSE19}, these  commits should always form a work item.  Other examples of real-world work items are logically related modifications of code, refactorings related to code cleaning tasks, and supplementary fixes spanning multiple commits, either because there were supplementary commits contributing to the same bug fix or because a bug fix was re-opened.  In this work, we focus on the subset of work items that include bug-introducing commits and their fixes.

For illustration, consider the code shown in \Cref{tab:code-example} from the dataset introduced by \cite{rosa21}.
Here,  the BIC and the FC insert statements into method \texttt{public static species make (...)} in \textit{source/RMG/jing/chem/Species.java}. Specifically, lines 880 to 888 were inserted in the BIC. It's worth noting that the last line added in the FC constitutes the actual bug fix.         
A traditional SZZ variant would identify the change in the FC and use Git blame to trace that fix back to a possible origin, where the bug was introduced.  However, in this instance, they would fail to work because the only changes to the lines of code in the FC involve newly added lines.  Thus, a Git blame has no history to trace back to, since those lines are new at the FC.
Traditional SZZ variants handle only code changes involving deletions and modifications to existing lines of code.
The design choice of \pfaszz to track method signature overlap between commits significantly enhances its capability to handle cases involving only code insertions in the FC. In the example under study, the code insertions occurred inside a method that was also modified in the BIC, thus revealing a common link that relates the FC and BIC. This link would not have been uncovered using line-level granularity approaches for cases involving only insertion code changes in the FC.

\section{Related Work}
\label{sec:related}

Recent research has focused extensively on fix-inducing changes, with the SZZ algorithm~\citep{szz} playing a crucial role in their identification. SZZ serves as the foundation for various studies aiming to understand defects better. For instance, seminal work and recent research~\citep{szz,eyolfson} have examined the timing of fix-inducing changes, highlighting patterns like occurrences on Fridays and between 12:00 AM and 4:00 AM. Additionally, SZZ has been instrumental in analyzing fix characteristics~\citep{pan09} and the duration of dormant bugs~\citep{dormant,snoring}. It's also been utilized to investigate the impact of organizational~\citep{authorship} and design choices on fix rates~\citep{clones}.

SZZ is central to just-in-time (JIT) defect prediction~\citep{jit}, aiming to forecast fix-inducing changes promptly upon proposal. Recent research~\citep{lapredict} questions the efficacy of advanced models, suggesting that a simple logistic regression model based on a single feature (lines added) consistently outperforms them. Nonetheless, newer studies indicate that with careful adjustments, deep learning approaches can surpass this baseline~\citep{ccrep}.

\begin{figure*}[ht]
\centering
\includegraphics[trim={3.7cm 12.6cm 4.6cm 1.8cm},clip,width=\linewidth]{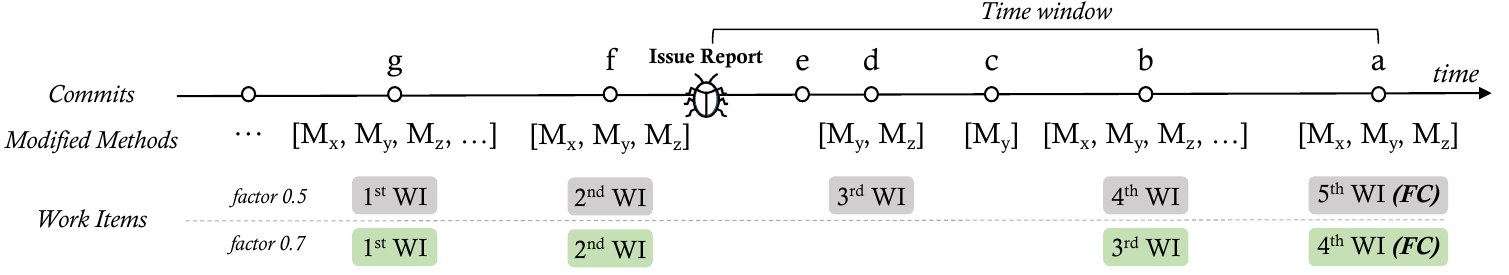}
\caption{Mining for work items. In general, the time window is a parameter specific to the algorithm using the discovered work items. For SZZ, we set the time window based on the dates of the fix commit (FC) and 30 days before the issue report.}
\label{fig:partial_fixes}
\end{figure*}

Given SZZ's importance, researchers have scrutinized its limitations. For instance, \cite{gema} found that SZZ often identifies code updated due to external factors, like API changes, rather than inherent bugs. Additionally, \cite{rezk22} identified "ghost commits" missed by SZZ. While noise in SZZ-labeled data has been extensively studied, recent research~\citep{fan19} suggests it has little impact on model performance.

To overcome limitations, the SZZ algorithm has undergone various enhancements. The original proposal~\citep{szz} employs a heuristic approach, filtering out implausible candidates. \cite{ag-szz} introduces AG-SZZ, utilizing an annotation graph for precise line tracing. \cite{l-szz} suggest L-SZZ and R-SZZ, improving the filtering stage. \cite{ma-szz} adds a stage for handling meta-changes (MA-SZZ). \cite{ra-szz,rastar-szz} integrate refactoring awareness (RA-SZZ, RA*-SZZ). Additionally, \cite{bludau2022pr} utilize pull request data for better mapping.

\cite{VSZZ_22} introduce V-SZZ, a modification of the SZZ algorithm targeting vulnerability-inducing commits in C/C++ and Java. However, V-SZZ's reliance on line mapping algorithms tailored to single-line code modifications may not effectively handle larger changes. Also, there is ambiguity in line mappings. For instance, applying AST-based line mappings in Java presents difficulties when lines at the fixing commit cannot be mapped to the previous commit, highlighting inaccuracies in mappings generated by GumTree (GT), MTDiff, or Iterative Java Matcher (IJM).  An example of this is the case when the lines were added in the FC and did not exist previously in the BIC. In contrast, our work focuses on identifying bug-inducing commits through a work item-based SZZ approach that tracks method overlap between the fix and bug-inducing commits. This allows it to track all changes, even if the FC change is adding new lines.

Neural SZZ~\citep{nszz} leverages a deep learning approach to rank deletion lines in a bug-fixing commit based on their likelihood of being the root cause of a bug. It models the semantic implications of deletion statements and their inter-relationships with other deletions and additions by constructing heterogeneous graphs. In these heterogeneous graphs, nodes represent deletion and addition statements; and edges their connections. Neural SZZ is limited to Java repositories because it relies on a specific static analyzer to build input graphs. In contrast, our approach works with any programming language. However, our method requires issue dates or a simulation of them, while their approach does not.  Like other SZZ variants, Neural-SZZ does not work with fixes that consist solely of insertions or deletions. On the contrary, \pfaszz is adept at handling such scenarios.

Previous research concentrated on improving SZZ algorithm performance metrics through various methods, such as mapping fix commits to bug-inducing ones or implementing additional candidate filtering. In contrast, our study is the first utilization of work items to pinpoint bug-inducing commits. Our approach utilizes a lightweight alternative to Git blame, identifying commits with logically related units of code changes. Moreover, as our method incorporates other SZZ variants when necessary, any enhancements made by those variants would also enhance our approach.

\section{Work Item Aware SZZ (\pfaszz)}
\label{sec:approach}

This section explains our work item-aware heuristic and its integration with SZZ (referred to as \pfaszz) to enhance SZZ's ability to detect BICs.

\subsection{Heuristic for Identifying Work Items}
\label{sec:wimotivation}

In our study of 818 experimental objects, at the 75\% percentile, the FC modifies 2 different methods, the BIC modifies 9 different methods, and the overlap of modified methods between FC and BIC is 1 method. Specifically, in 667 out of the 818 objects (81.54\%), the BIC and FC modified at least one common method.
Additionally, \cite{pan_09} demonstrated that many bug fix features occur within method bodies or declarations.
Our heuristic utilizes these insights by tracking changed methods from the fix commit over time to identify work items that lead to the BIC.

\Cref{fig:partial_fixes} illustrates our heuristic-based approach to identifying commits belonging to a work item. In the first step, we use \emph{PyDriller}~\citep{pydriller} to identify the modified (including deleted) methods in the input commit (in this paper, that is always FC).  In the second step, we walk backward to older commits and build a tracking matrix based on the set of modified methods in that commit, assigning a unique ID to each modified method.

\begin{table}[ht]
\caption{Tracking matrix for \Cref{fig:partial_fixes}. Work items $W\!\!I(i)$ are computed using \Cref{eq:filtering_factors}.}
\label{tab:matrix}
\def\arraystretch{1.6}
\setlength\tabcolsep{5pt}
\setlength\arrayrulewidth{0.7pt}
\small
\centering
\begin{tabular}{|c|c|c|c|@{}p{2pt}@{}|c|c|}
\hhline{|----|~|--|}
\multicolumn{4}{|c|}{\textbf{Tracking Matrix}} & & \multicolumn{2}{c|}{\bf \em WI(i)} \\
\hhline{|----|~|--|}
\textbf{Commit (i)} & $\mathbf{M_x}$ & $\mathbf{M_y}$ & $\mathbf{M_z}$ & & \textbf{factor = 0.5} & \textbf{factor = 0.7} \\
\hhline{|----|~|--|}
a & \textbf{1} & \textbf{1} & \textbf{1} & & \cellcolor{black!20}True & \cellcolor{lightgreen}True \\
\hhline{|----|~|--|}
b & \textbf{1} & \textbf{1} & \textbf{1} & & \cellcolor{black!20}True & \cellcolor{lightgreen}True \\
\hhline{|----|~|--|}
c & 0 & \textbf{1} & 0 & & False & False \\
\hhline{|----|~|--|}
d & 0 & \textbf{1} & \textbf{1} & & \cellcolor{black!20}True & False \\
\hhline{|----|~|--|}
e & 0 & 0 & 0 & & False & False \\
\hhline{|----|~|--|}
f & \textbf{1} & \textbf{1} & \textbf{1} & & \cellcolor{black!20}True & \cellcolor{lightgreen}True \\
\hhline{|----|~|--|}
g & \textbf{1} & \textbf{1} & \textbf{1} & & \cellcolor{black!20}True & \cellcolor{lightgreen}True \\
\hhline{|----|~|--|}
\end{tabular}
\end{table}

For example, as shown in \Cref{tab:matrix}, the input commit \texttt{a} modified three methods: $M_x$, $M_y$, and $M_z$, as indicated by ones in those three columns. The prior commit \texttt{b} also modifies the same three methods, as well as other methods that were not modified in \texttt{a}.  These other modified methods are ignored in the tracking matrix. Other commits, such as commit \texttt{c}, only modified one method, $M_y$, resulting in a single one across its matrix row.

In the final step, we apply a filter to identify if a commit is a work item, based on the number of modified methods.  We describe our function $W\!I(i)$ in \Cref{eq:filtering_factors} for identifying if a commit in row $i$ is a work item:

\begin{equation}
W\!I(i) = \sum_{j \in \{x,y,z\}} T_{i,j} \ge \text{factor} \times N
\label{eq:filtering_factors}  
\end{equation}

\noindent where, $T$ is the tracking matrix, $i$ denotes the row for a commit id (e.g., $a,\ldots,g$) and $j$ denotes the column for a method id (e.g., $x,y,z$).  $factor$ is a decimal value between 0 and 1. In our study, we evaluate factor values from $0.3$ to $0.9$ and choose $0.7$ as the optimal value (see \Cref{sec:workitemstats}).  $N=|\{M_x,M_y,M_z\}|$ is the number of modified methods in the input commit FC or number of columns.

For example, consider commit \texttt{d} in \Cref{fig:partial_fixes}, which modifies methods $M_y$ and $M_z$.  In this example, the number of tracked methods (\emph{N}) is 3. If we set the factor to 0.5, the number of modified methods (2 in this case) is greater than or equal to 0.5 $\times$ 3 (or 1.5). Therefore, we mark ``True'' in the matrix when the factor is 0.5.  On the other hand, if we set the factor to 0.7, the number of modified methods is not greater than or equal to 0.7 $\times$ 3 (or 2.1).  Therefore, we mark ``False'' in the matrix when the factor is 0.7.  The choice of what factor to filter by is a heuristic parameter.  We later discuss the concrete factor used in this paper and how we choose it.

\subsection{Enhancing SZZ with Work Item Awareness}
\label{subsec:wiaszz}

We introduce \pfaszz, a new variant of the SZZ algorithm. This variant utilizes our work item finding heuristic to identify related commits to the FC.  Our hypothesis is that work items are often also BICs, leading to a more accurate identification of BICs than other SZZ variants. \Cref{fig:partial_fixes} illustrates the process that \pfaszz, uses to select \emph{a single BIC candidate from a list of work items}. First, \pfaszz evaluates the distribution of work items within a time window between the issue report date and FC (i.e., commit \texttt{a} in this example). As shown, our work items include commits \texttt{a}, \texttt{b}, \texttt{d}, \texttt{f}, and \texttt{g} when we use a factor of 0.5 and commits \texttt{a}, \texttt{b}, \texttt{f}, and \texttt{g} for a factor of 0.7. If there is at least one work item before the time window (i.e., commits \texttt{f} and \texttt{g}), \pfaszz suggests the commit closest to the issue report date as the BIC (i.e., commit \texttt{f}). However, if there are no work items before the issue report date, \pfaszz simply invokes a default SZZ variant to produce BIC candidates.

Given that the heuristic for finding work items in \pfaszz encompasses a broader and more general approach of solely pinpointing the work item commit aligned with the BIC annotated for a specific FC, many of the identified work items from an FC may not precisely match the annotated BIC but still pertain to the FC. This reasoning led us to opt for our predicted BIC as the singular work item closest to the bug report date. This choice stems from the assumption that the commit introducing a bug typically occurs around the time it is observed and reported. Variants of SZZ, such as L and R, also adhere to a single predicted BIC framework.

In summary, \pfaszz employs two approaches to select BIC candidates. The first approach is based on work items. It works with any repository instance having work items that exist before the issue report date and selects only the newest work item before the issue report date as the BIC candidate. The second approach is for any repository instance without a work item before the issue report date, where it falls back to pre-existing SZZ implementations to select BIC candidates. In the next sections, we describe our evaluation methodology and report the results of our study.

\section{Empirical Evaluation}
\label{sec:eval}

In this study, we consider the following research questions:

\begin{rqs}
    \item\label{rq:manual} \textbf{Does our heuristic find work items?}  Are the identified commits work items that could be logically grouped together?

    \item\label{rq:compare} \textbf{Does our work item heuristic help increase the accuracy of SZZ?}  How does \pfaszz perform compared to other SZZ variants in terms of precision, recall, and $F_1$ on a previously created oracle?

    \item\label{rq:filter} \textbf{Does applying an issue-date filter improve the accuracy of SZZ and/or \pfaszz?}  A prior study showed that an issue-date filter can improve the precision of SZZ. Do we see similar results with our dataset?  Does a similar filter also help improve \pfaszz?

    \item\label{rq:overfit} \textbf{Does always selecting the candidate nearest the issue date cause overfitting?}  Since our approach relies on a filter that always selects exactly one candidate BIC based on the issue date, that approach might overfit the evaluation dataset.  Here we investigate if that is the case or not.
\end{rqs}

In the next section, we describe the experimental objects used to answer these research questions.

\subsection{Experimental Data}
\label{sec:experimentaldata}

We utilized the dataset from \cite{rosa21}, comprising a total of 1,115 Fix Commits (FCs) across eight programming languages (C, Python, C++, JavaScript, Java, PHP, Ruby, and C\#). Each fix commit (FC) is annotated with its originating repository information, the programming language(s) it employs, the Bug-Introducing Change (BIC) it fixes, and the date of the bug report. Additionally, we used \emph{PyDriller}~\citep{pydriller} to track modified methods and files across the commit history associated with each FC.
For this study, we focused on a subset of 818 FCs, as outlined in \Cref{tab:fcs}.  These were what remained from the original 1,115 FCs after removing unresolved commit hashes, commits with no modified (non-test) methods, items with clearly identified wrong issue dates, and items with manually identified wrong BICs indicated.

For \ref*{rq:manual} we then sample that down to 256 work items. For \ref*{rq:compare}--\ref*{rq:overfit}, we partition the 818 FCs into two parts and focus on the 421 that generated work items. We interchangeably refer to this set of FCs as our experimental \textit{objects} going forward.

\begin{table}[ht]
    \centering
    \caption{Detail of removed FCs out of the 1,115 FCs in the dataset by \cite{rosa21}.}
    \label{tab:fcs}
    \begin{tabular}{l|c}
      \textbf{Reason}                                  & \textbf{Counts} \\
      \midrule
      Unresolved commit hashes in FC's commit history  & 127 \\
      No modified methods in FC (except tests)         & 149 \\
      Wrong issue dates (post-FC) by manual evaluation & 18  \\
      Wrong BIC by manual evaluation                   & 3   \\
      \bottomrule
      Total Removed FCs                                & 297 \\
      Final FCs (Experimental Objects) & 818  \\
 \end{tabular}
 \end{table}

\subsubsection{Issue Date Offsets}
\label{sec:issueoffsets}

We found that out of 818 experimental objects, only 91 had actual issue report dates, crucial for BIC candidate selection. For the remaining 727 objects, simulated dates were set to 60 seconds after the BIC, following an approach used by \cite{rosa21}.

Further examination revealed discrepancies in FC-BIC time windows, ranging from 1 day to 10 years. To address this, we propose computing issue report dates based on offsets derived from real instances (see \Cref{eq:newbugreport}).

\begin{equation}
\centering
\text{issue\_offset($i$)}=\frac{\mbox{IssueReportDate}_i - BIC_i}{FC_i-BIC_i}
\label{eq:newbugreport}
\end{equation}

Based on the 91 issues, the median offset (Q2) of 93\% was chosen for its practicality and robustness. We also evaluated the effects of various offset values (Q1 and Q3) showed sensitivity but no bias toward Q2. Choosing Q1 increased \pfaszz performance by 1.9\%, while Q3 decreased it by 0.9\%. However, Q2 remained the optimal choice for balance.

Note that only one set of issue report dates was used for analysis: the original 91 with real dates and the 727 with Q2 offsets, ensuring consistency across evaluations.

\subsubsection{Selecting the Filtering Factor}
\label{sec:workitemstats}

\Cref{tab:pfaszz} presents the findings from a sensitivity analysis conducted on the filtering factor proposed in \Cref{eq:filtering_factors} for identifying work items. The table comprises eight columns. The first column denotes the utilized factor, while the second column indicates the count of experimental objects (FCs) with work items occurring before the issue date, which varied depending on the applied factor. The third column displays the number of work items (WIs) generated by the respective objects for each factor. The final four columns depict the performance metrics of \pfaszz concerning FCs, as our method aims to select a single work item for BIC. Additionally, note that the sum of TP and FP for each row corresponds to the FC count within that row.

\begin{table}[ht]
\centering
\caption{\pfaszz results for different factors. The second column is the number of FCs or experimental objects $n$ (out of 818) that produced work items at that factor.}
\label{tab:pfaszz}
\rowcolors{2}{white}{gray!10}
\begin{tabular}{crrrrrrr}
  \toprule
  \multicolumn{1}{c}{\textbf{Factor}} & \multicolumn{1}{c}{\textbf{FCs($n$)}} & \multicolumn{1}{c}{\textbf{WIs}} & \multicolumn{1}{c}{\textbf{TP}} & \multicolumn{1}{c}{\textbf{FP}} & \multicolumn{1}{c}{\textbf{Recall}} & \multicolumn{1}{c}{\textbf{Precision}} & \multicolumn{1}{c}{$\boldsymbol{F_1}$} \\
  \midrule
  \textbf{0.3} & 512 & 1068 &375 & 137 & 0.733 & 0.732 & 0.733 \\
  \textbf{0.5} & 483 &  979 &360 & 123 & 0.748 & 0.745 & 0.746 \\
  \highlight{0.7} & 421  & 754 & 325 & 96 & \highlight{0.773} & \highlight{0.773} & \highlight{0.773} \\
  \textbf{0.9} & 408 & 739 &314 & 94  & 0.770 & 0.770 & 0.770 \\
  \bottomrule
  \end{tabular}     
  \end{table}

Experimenting with factors of 0.3, 0.5, 0.7, and 0.9 (\Cref{tab:pfaszz}), we found that by varying the filtering factor from 0.3 to 0.9, our heuristic generated work items for objects that decrease from 512 to 408. Note that the objects under the $FCs$ column in \Cref{tab:pfaszz} produced work items with at least one work item before the issue date.
The objects without work items lacked modified methods at the FC or had no overlapping methods in the commit history, as detailed in \Cref{sec:threats}.

From \Cref{tab:pfaszz}, we conclude that the best-performing factor is 0.7, based on the highest obtained $F_1$-score of 0.77. We utilize this factor throughout our evaluation, meaning we used 421 objects with work items for running the part of \pfaszz that processes work items. The breakdown of objects used for all parts of the process is also depicted graphically in \Cref{fig:overview} for the best-performing factor of 0.7.

Since the best-performing factor of 0.7 was based on a large number of sampled projects, we believe it is a reasonable default value for the parameter.  Future projects would most likely start with the value set to 0.7 and then possibly fine-tune it on their own project's data if they are not seeing reasonable results.

\subsection{Study Operation}

We conducted our experiments on a desktop computer with an Intel Core i7-10700 CPU and 32GB of RAM running Ubuntu Linux 18.04.6 LTS.  We implemented \pfaszz and our data collection and processing tools in Python.  We used the original implementations for SZZ variants from \cite{rosa21} (that requires version 1.15.2 of \emph{PyDriller}).  For our heuristic implementation, we used a later version of \emph{PyDriller} 2.4.1 to identify modified files and methods.

Concerning \ref*{rq:manual}, we used three authors as raters. The first two are Computer Science Ph.D. students with 12 and 7 years of programming experience. The last rater is a software developer with over two decades of experience.

\subsection{Variables and Measures}
\label{sec:vars_methods}
Here we outline our study's (in)dependent variables.

\subsubsection{Dependent Variables}
\label{subsec:depvars}

As dependent variables, we chose metrics allowing us to answer each of our research questions.

For \ref*{rq:manual}, the evaluation aims to determine if a \emph{predicted work item and its corresponding FC are logically related and should be considered a single unit of work}.  Three answers are:

\begin{enumerate}
    \item \textbf{YES}: the predicted work item is related to FC
    \item \textbf{NO}: the predicted work item is not related to FC
    \item \textbf{UNKNOWN (UNK)}: a relation could not be determined
\end{enumerate}

We quantified agreement among raters using \textbf{Cohen's \emph{Kappa}} and \textbf{Gwet's \emph{AC1}} values. Both approaches measure the agreements between two raters; however, the two approaches use two distinct comparators.  Kappa compares the number of observed agreements with the number of expected agreements.  AC1, on the other hand, compares the number of observed agreements with the number of expected disagreements~\citep{gwet08:_comput}. 

We measured \textbf{Accuracy} as the ratio between actual work items manually verified to be related to its FC and the total number of predicted work items:

\begin{equation}
\centering
\text{Accuracy}=\frac{|WI_{\text{verified}}|}{|WI_{\text{total}}|}
\end{equation}

For \ref*{rq:compare}-\ref*{rq:overfit}, the variables below are used in the performance analysis of \pfaszz and SZZ variants.

\noindent
\textbf{True Positive (TP)} is the number of BIC candidates that have been \emph{correctly predicted} by each approach across all experimental objects. 

\noindent
\textbf{False Positive (FP)} is the number of BIC candidates that have been \emph{incorrectly predicted} by each approach across all experimental objects. 

\noindent
\textbf{Recall} is the ratio between the correctly predicted BICs and all BICs in the oracle:
\begin{equation}
\centering
\text{Recall}=\frac{|TPs|}{|BICs|}
\end{equation}

\noindent
\textbf{Precision} is the ratio of TPs to the sum of TPs and FPs:
\begin{equation}
\centering
\text{Precision}=\frac{|TPs|}{|TPs|+|FPs|}
\end{equation}

\noindent
\textbf{$F_1$-score} is the harmonic mean between precision and recall:
\begin{equation}
\centering
\text{$F_1$}=2*\frac{Precision*Recall}{Precision+Recall}
\end{equation}

\subsubsection{Independent Variables}
Our independent variables involve the algorithms used in our study. We consider \pfaszz and six variations of SZZ (B, AG, L, R, MA, and RA for Java) implemented by \cite{rosa21}. We did not consider Neural-SZZ as, despite assistance from the authors, we were unable to obtain the trained models or recreate them ourselves to reproduce the results presented in the paper. Note that \pfaszz always reports exactly one BIC candidate for each input FC, while the other SZZ variants report zero or more candidates per input FC.

For the SZZ variants, we also applied \issuefilter, a post-processing filter that removes any BIC false positive candidates occurring after the issue report date.

\section{Results and Analysis of \pfaszz}

In this section, we provide the results of our evaluation of our work item finding heuristic and SZZ variant \pfaszz.

\subsection{\ref*{rq:manual}: Does our heuristic find work items?}
\label{sec:man_eval}

To begin, we created a substantial sample to conduct a manual evaluation of the work items produced by our most effective factor, set at 0.7.  With this factor, there were 421 objects (FCs) generating a total of 754 work items. We then extracted a statistically significant sample of 256 work items for manual assessment, ensuring a 95\% confidence level with a 5\% margin of error. This sample comprised 123 instances of BIC and 133 instances of non-BIC, addressing any skewed distributions by selecting every third non-BIC work item per FC.

Raters determined whether each work item was logically related to its FC and should be considered a single work unit. The three possible responses to categorize the relatedness were YES, NO, or UNK.

Notably, raters did not establish specific rules for associating work items. Instead, they followed a general, open guideline to determine if a given commit and its FC are a work item by answering the question, `Should these commits be ideally merged into a single commit?'. Raters had access to each commit's GitHub URL, allowing them to view files changed, diffs, source code changes, tags, parent commit, author, committer, date, time, and commit message log (except for any references to the BIC found in it). To filter any references to BIC, raters used a browser plugin they developed to hide these commit references. This step was necessary to prevent bias when evaluating work items that matched BIC commits, given the dataset's construction in \cite{rosa21}, which links each FC with the BIC it is fixing, and also ensured that judgments of current commit changes remained unbiased and independent of any judgment that raters could derive from looking at the code changes in the BIC.

\begin{table}[ht]
\centering
\caption{Manual evaluation of the heuristic for identifying work items.\\``All WIs" refers to all work items generated from the FCs. ``Only BIC" refers to those work items that match a BIC. Values are the agreements reached after joint discussion by the first two raters once each batch was individually evaluated.}
\label{tab:man_eval}
\rowcolors{3}{white}{gray!10}
\renewcommand{\tabcolsep}{0.7em}
\begin{tabular}{rrrr|rrrrr}
  \toprule
  \multicolumn{1}{c}{} & \multicolumn{3}{c|}{\textbf{Only BIC}} & \multicolumn{5}{c}{\textbf{All Work Items}} \\
  \multicolumn{1}{c}{} & \textbf{YES} & \textbf{NO} & \textbf{UNK} & \textbf{YES} & \textbf{NO} & \textbf{UNK} & \textbf{Kappa} & \textbf{AC1} \\
  \cmidrule(l{0.5em}r{0.5em}){2-4}\cmidrule(l{0.5em}r{0.5em}){5-7}\cmidrule(l{0.5em}r{0.5em}){8-9}
  \textbf{1}   &  24 & 0 & 0 &  38 & 12 & 0 & 0.329 & 0.583 \\
  \textbf{2}   &  26 & 1 & 0 &  36 & 20 & 0 & 0.542 & 0.690 \\
  \textbf{3}   &  24 & 0 & 0 &  29 & 21 & 0 & 0.493 & 0.650 \\
  \textbf{4}   &  24 & 0 & 0 &  31 & 19 & 0 & 0.425 & 0.597 \\
  \textbf{5}   &  24 & 0 & 0 &  33 & 17 & 0 & 0.651 & 0.793 \\
  \midrule
  \textbf{All} & 122 & 1 & 0 & 167 & 89 & 0 & 0.492 & 0.662 \\
  \bottomrule
  \end{tabular}\end{table}

After rating each batch, the raters met to discuss and reach a consensus. By discussing after each batch, their collective experience helped calibrate the processes for the next batches. 
\Cref{tab:man_eval} shows in columns 5-7 the results of each batch after discussions; and the agreement metrics for the overall evaluation in columns 8-9.

\subsubsection{Inter-rater Reliability Metrics}
\label{sec:irr-metrics}

We assessed agreement using Cohen's kappa, recommended by the ACM SIGSOFT Empirical Standards Supplement on Inter-Rater Reliability and Agreement~\citep{ralph_etal20:_empir_stand_softw_engin_resear}, and Gwet's AC1~\citep{gwet08:_comput}. Kappa measures observed versus expected agreements, while AC1 compares observed agreements to expected disagreements~\citep{gwet08:_comput}.
These approaches produce orthogonal results~\citep{vach_23}. Kappa can be affected by the base-rate paradox, potentially lowering its value below observed agreements, unlike AC1. In our evaluation (see Table 1), kappa indicates moderate agreement (0.492), while AC1 shows stronger agreement (0.662).
Notably, focusing on instances labeled as BICs, raters disagreed on only one instance out of 123 instances, which was later confirmed as mislabeled. This reaffirms BICs as work items, as suggested by~\citep{mcintosh11:_work_items}.

\subsubsection{Non-BIC Evaluation Process}

In our manual review, 133 work items from an FC didn't match the provided BIC commit. We termed these as non-BIC instances. To further assess this subset, we involved a third rater with over two decades of programming experience, who had access to the same information as the other raters but without the injected JavaScript and with visibility of the BIC links for each instance.

\begin{table}[ht]
\centering
\caption{Manual Evaluation of Heuristic for Identifying Work Items by 3 Raters.\\``BIC" denotes work items that align with the oracle BIC from the original dataset for each FC. ``non-BIC" indicates work items not matching a BIC but generated from each FC and selected for the evaluation sample.}
\label{tab:maneval_accuracy}
\rowcolors{2}{white}{gray!10}
\begin{tabular}{lrr|rr}
  \toprule
   & \textbf{BIC} & \textbf{non-BIC} & \multicolumn{2}{c}{\textbf{Total}} \\
  \midrule
  \textbf{YES} & 122 & 40 & 162 & (64.8\%) \\
  \textbf{NO} & 1 & 87 & 88 & (35.2\%) \\
  \textbf{UNK} & 0 & 6 & 6 & \multicolumn{1}{c}{--} \\
  \bottomrule
  \end{tabular}
\end{table}

\begin{table*}[ht!]
\centering
\caption{Comparison of SZZ variants and \pfaszz.  For all deltas except FP, higher is better. The first table is for TP and FP, and the second is for Recall, Precision and F1.}
\label{tab:all_pfaszzdeltas}
\rowcolors{3}{white}{gray!10}
\renewcommand{\tabcolsep}{0.44em}
\resizebox{0.75\textwidth}{!}{
\begin{tabular}{lrrrrrrrr}
\toprule
 &  & \multicolumn{3}{c}{\textbf{TP}} & \multicolumn{3}{c}{\textbf{FP}} \\
 & \textbf{n} & \textbf{SZZ} & \textbf{\pfaszz} & \textbf{$\Delta$} & \textbf{SZZ} & \textbf{\pfaszz} & \textbf{$\Delta$} \\
\cmidrule(l{0.5em}r{0.5em}){3-5}\cmidrule(l{0.5em}r{0.5em}){6-8}
\textbf{B} & 818 & 585 & 575 & -10 & 939 & 808 & -131 \\
\textbf{AG} & 818 & 482 & 510 & 28 & 461 & 395 & -66 \\
\textbf{L} & 818 & 351 & 432 & 81 & 269 & 260 & -9 \\
\textbf{R} & 818 & 442 & 486 & 44 & 171 & 201 & 30 \\
\textbf{MA} & 818 & 481 & 513 & 32 & 549 & 465 & -84 \\
\textbf{RA} & 59 & 25 & 30 & 5 & 29 & 30 & 1 \\
\midrule
\textbf{mean} &  &  &  & +30 &  &  & -43.17 \\
\textbf{median} &  &  &  & +30 &  &  & -37.40 \\
\bottomrule
\end{tabular}
}
\end{table*}
\begin{table*}[ht!]
\centering
\rowcolors{3}{white}{gray!10}
\renewcommand{\tabcolsep}{0.44em}
\resizebox{\textwidth}{!}{
\begin{tabular}{lrrrrrrrrrrrr}
\toprule
 &  & \multicolumn{3}{c}{\textbf{Recall}} & \multicolumn{3}{c}{\textbf{Precision}} & \multicolumn{3}{c}{$\boldsymbol{F_1}$} \\
 & \textbf{n} & \textbf{SZZ} & \textbf{\pfaszz} & \textbf{$\Delta$} & \textbf{SZZ} & \textbf{\pfaszz} & \textbf{$\Delta$} & \textbf{SZZ} & \textbf{\pfaszz} & \textbf{$\Delta$} \\
\cmidrule(l{0.5em}r{0.5em}){3-5}\cmidrule(l{0.5em}r{0.5em}){6-8}\cmidrule(l{0.5em}){9-11}
\textbf{B} & 818 & 0.71 & 0.70 & -0.01 & 0.38 & 0.41 & 0.03 & 0.50 & 0.52 & 0.02 \\
\textbf{AG} & 818 & 0.59 & 0.62 & 0.03 & 0.51 & 0.56 & 0.05 & 0.54 & 0.59 & 0.05 \\
\textbf{L} & 818 & 0.43 & 0.53 & 0.10 & 0.56 & 0.62 & 0.06 & 0.48 & 0.57 & 0.09 \\
\textbf{R} & 818 & 0.54 & 0.59 & 0.05 & 0.72 & 0.71 & -0.01 & 0.62 & \textbf{0.64} & 0.02 \\
\textbf{MA} & 818 & 0.59 & 0.63 & 0.04 & 0.46 & 0.52 & 0.06 & 0.52 & 0.57 & 0.05 \\
\textbf{RA} & 59 & 0.42 & 0.50 & 0.08 & 0.46 & 0.50 & 0.04 & 0.44 & 0.50 & 0.06 \\
\midrule
\textbf{mean} &  &  &  & +0.05 &  &  & +0.04 &  &  & +0.05 \\
\textbf{median} &  &  &  & +0.04 &  &  & +0.05 &  &  & +0.05 \\
\bottomrule
\end{tabular}
}
\end{table*}

After discussion, as shown in \Cref{tab:maneval_accuracy}, all three raters agreed on 40 YES ratings.  In other words, even though these commits are not BICs, they are still logically related and should be considered a single unit of work (ideally merged into a single commit).  There were 6 UNK ratings, indicating it was difficult to determine if a relationship existed.

In total, there were 43 disagreements (32\%) between the evaluation results of the first two raters and the third rater for the non-BIC judgments. Cohen's kappa statistic is 0.275 (fair level of agreement) and the AC1 value is 0.585.

\subsubsection{Manual Evaluation and Heuristic Discussion}
\label{subsec:maneval_discussion}

While the kappa and AC1 values quantify the agreement between our three raters, the ultimate metric to determine the effectiveness of our heuristic is its accuracy in selecting work items. Based on \Cref{tab:maneval_accuracy}, our heuristic detected 162 work items. This yields an accuracy of at least 63.28\% (162/256). It is worth noting the heuristic found 122 BICs out of 123, and the one not found turned out to be an incorrect oracle.
These results show evidence that the concept of work items can be effectively used to find BICs. 

Note that the non-BIC instances present more challenges for our heuristic. Out of 133 instances, we only verified 40 work items related to its FC (see \Cref{tab:maneval_accuracy}). There were also six that we could not manually determine if they were work items due to large modifications involving multiple files.

Furthermore, we also observed a few objects (3175, 2399, 1582) where the oracle BICs from \cite{rosa21} were incorrect. In object 3175, the actual BIC extracted code into a function, leading to a variable assignment not being sent back to the call site due to function scoping. However, the developer incorrectly identified the commit hash when fixing the issue, resulting in an incorrect BIC.
In object 2399, we found through manual inspection that the oracle BIC updates a version number in a setup file, whereas the FC modifies an unrelated else branch in a function located in a different file from the setup file.
Finally, in object 1582, we found an incorrect BIC. Both the FC and a mined work item modify the same function in the same file, but the oracle BIC modifies a different file. Thus, we suggest the work item related to the fix as a more accurate BIC.

\subsubsection{Threats to Manual Evaluation Validity}
\label{sec:thre-manu-eval}

The rating process may introduce bias, especially since all raters are co-authors of the paper. However, this potential bias is mitigated by several factors. First, apart from the periods during which the first two raters discussed each batch, there was no discussion of the rating process, enhancing the independence of ratings and reducing bias. Second, the fact that all three raters have different educational backgrounds further contributes to bias reduction. Third, the addition of a third rater with access to the BIC helps improve the accuracy. Fourth, the three authors had significantly different roles in conceiving this paper, with one primarily serving as an additional rater.

\findings{
Our heuristic accurately identifies work items that match BIC labels with a success rate of 122 out of 123, which validates that bug-inducing and fixing changes should be considered a single unit of work. Overall, it has a 64.8\% accuracy in detecting work items.
}

\subsection{\ref*{rq:compare}: Does our work item heuristic help increase the accuracy of SZZ?}

We tested all six SZZ variants and \pfaszz on 818 objects. However, the refactoring-aware SZZ (RA) is specific to Java and was limited to 59 repositories. We universally applied the \issuefilter, which is known to reduce false positives and increase precision. 
In \Cref{tab:all_pfaszzdeltas}, we detail true positives (TP), false positives (FP), recall, precision, $F_1$, and their differences ($\Delta$). The ``SZZ'' columns depict results solely from SZZ, while the "\pfaszz" columns show our approach. Our method suggests the BIC before the time window in \Cref{fig:partial_fixes}; otherwise, it calls SZZ. With a set factor of 0.7, 421 objects detected work items.
Comparing true positives, \pfaszz excels across all variants except the baseline (B), yielding up to 81 more BICs. On average, we find nearly 31 more BICs than SZZ variants, with the highest recall boost of 11\% seen in the best case (L).
Regarding false positives, \pfaszz outperforms all but (R) and (RA) variants, notably reducing false positives by 131 in the best case (B). On average, it decreases false positives by 43 compared to SZZ variants, leading to a precision gain ranging from 3\% to 6\%.

In summary, \pfaszz detects more true BICs while reducing false positives. Excluding the baseline, it consistently enhances true positive detection while decreasing false positives, resulting in higher $F_1$ scores beating all six variants, with an average increase of 5\%.

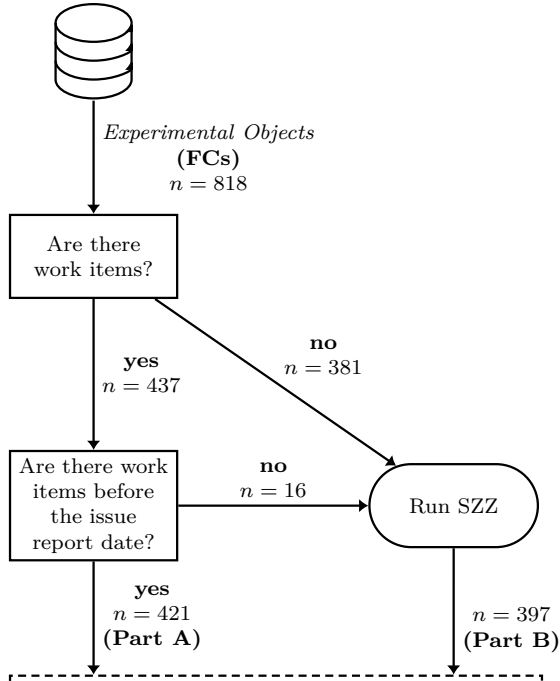
\begin{figure}[t]
  \centering
  \makeatletter
  \tikzset{
      database/.style={
          path picture={
              \draw (0, 1.5*\database@segmentheight) circle [x radius=\database@radius,y radius=\database@aspectratio*\database@radius];
              \draw (-\database@radius, 0.5*\database@segmentheight) arc [start angle=180,end angle=360,x radius=\database@radius, y radius=\database@aspectratio*\database@radius];
              \draw (-\database@radius,-0.5*\database@segmentheight) arc [start angle=180,end angle=360,x radius=\database@radius, y radius=\database@aspectratio*\database@radius];
              \draw (-\database@radius,1.5*\database@segmentheight) -- ++(0,-3*\database@segmentheight) arc [start angle=180,end angle=360,x radius=\database@radius, y radius=\database@aspectratio*\database@radius] -- ++(0,3*\database@segmentheight);
          },
          minimum width=2*\database@radius + \pgflinewidth,
          minimum height=3*\database@segmentheight + 2*\database@aspectratio*\database@radius + \pgflinewidth,
      },
      database segment height/.store in=\database@segmentheight,
      database radius/.store in=\database@radius,
      database aspect ratio/.store in=\database@aspectratio,
      database segment height=0.25cm,
      database radius=0.5cm,
      database aspect ratio=0.5,
  }
  \makeatother
  \begin{tikzpicture}[node distance=1.5cm and 2.5cm,
    decision/.style={draw,line width=0.3mm,font=\small,text width=2cm,minimum height=1.1cm,align=center},
    every path/.style={-{Latex[width=1.5ex,length=1ex]},draw,line width=0.3mm},
    every edge quotes/.style={auto,align=center,font=\footnotesize}]
    \node (input) [database] {};
    \begin{scope}[nodes=decision]
      \node (start) [below=of input] {Are there work items?};
      \node (window) [below=2cm of start] {Are there work items before the issue report date?};
      \node (SZZ) [rounded rectangle,right=of window] {Run SZZ};
    \end{scope}
    \node (output) [draw,densely dashed,text width=6.8cm,below=of window.south west,anchor=north west,align=center] {\textit{bug-inducing commits}};
  
    \draw (input) edge["\textit{Experimental Objects}\\\textbf{(FCs)}\\\(n=818\)"] (start);
    \draw (start) edge["\textbf{yes}\\\(n=437\)"] (window);
    \draw (start) edge["\textbf{no}\\\(n=381\)"] (SZZ);
    \draw (window) edge["\textbf{no}\\\(n=16\)"] (SZZ);
    \draw (SZZ) edge["~\\~\\\(n=397\)\\\textbf{(Part B)}"] (output.north -| SZZ);
    \draw (window) edge["\textbf{yes}\\\(n=421\)\\\textbf{(Part A)}"] (output.north -| window);
  \end{tikzpicture}
  \vspace{-2em}
   \caption{Overview of the \pfaszz algorithm. An instance with at least one work item before the issue report date is processed in Part A using our work item heuristic, otherwise the underlying SZZ processes the instance in Part B.}
  \label{fig:overview}
\end{figure}

In \Cref{fig:overview} we show a breakdown of how \pfaszz partitions the data into two parts: Part A, where it was able to detect a work item grouping and Part B where it detected no work items and thus falls back to just calling the original SZZ algorithm.  Next, we break down the results into these two parts, showing Part A in \Cref{tab:all_szz} and Part B in \Cref{tab:all_pfaszz}.

\subsubsection{Part A: FCs With Work Items}
\label{subsec:partA}

We observe that Part A generally has more true positives and fewer false positives than Part B.  It seems even the SZZ variants perform better on this subset of the data.  That said, \pfaszz still outperforms the SZZ variants by simultaneously maintaining a high recall and precision rate.  On this subset of the data, our approach outperforms the SZZ variants with a higher $F_1$ ranging from +3\% to +14\%.

As mentioned previously in \Cref{sec:bg}, existing SZZ variants do not handle cases where fixes only involve code insertions, and few of them work when code changes are only deletions. Further examination of our experimental data reveals that 71 out of 818 objects fall into this category (i.e., 9 objects with only deletions and 62 objects with only insertions). Among these, \pfaszz processed 35 objects, accounting for 49.30\%—comprising 8 deletion-only objects and 27 insertion-only objects. \pfaszz successfully identified 23 BICs, representing a rate of 65.71\%.

Moreover, we assessed the effectiveness of our approach when experimental objects include actual issue report dates. Among the 818 objects, 91 had issue report dates. Of these, 43 were processed by Part A and 48 by Part B. For the 43 processed by \pfaszz, it achieved a precision, recall, and F1 score of 74\%. This underscores the effectiveness of \pfaszz when issue report dates are provided.

\begin{table}[t]
\centering
\caption{\pfaszz and SZZ variants performance (Part A). Note: \pfaszz only reports one row as it never runs SZZ on these inputs.}
\label{tab:all_szz}
\rowcolors{3}{white}{gray!10}
\newcommand{\oldtabcolsepaaaa}{\tabcolsep}
\renewcommand{\tabcolsep}{0.67em}
\begin{tabular}{l@{\hskip0.5cm}rrrrrrr}
\toprule
 & \multicolumn{7}{c}{\textbf{Part A (All Languages)}} \\
 & \textbf{n} & \textbf{TP} & \textbf{FP} & \textbf{Recall} & \textbf{Precision} & $\boldsymbol{F_1}$ & \textbf{$\boldsymbol{F_1} \Delta$} \\
\cmidrule(l{-0.5em}r{0em}){2-8}
\textbf{WIA-SZZ} & 421 & 325 & 95 &\textbf{0.77} & \textbf{0.77} & \highlight{0.77} & -- \\
\textbf{B} & 421  &\highlight{335}  &226  &0.60       &\highlight{0.80}  &0.68 & +0.09 \\
\textbf{AG} & 421  &297  &161  &0.65       &0.71  &0.68 & +0.09  \\
\textbf{L} & 421  &244  &104  &0.70       &0.58  &0.63 & +0.14  \\
\textbf{R} & 421  &281  &\highlight{65}   &\highlight{0.81}       &0.67  &0.74 & +0.03 \\
\textbf{MA} & 421  &293  &179  &0.62       &0.70  &0.66 & +0.11  \\
\rowcolor{white}\\
 & \multicolumn{7}{c}{\textbf{Part A (Java)}} \\
\cmidrule(l{-0.5em}r{0em}){2-8}
\textbf{WIA-SZZ} & 29 & \highlight{22} &7 & \highlight{0.76} & \highlight{0.76} & \highlight{0.76} \\
\textbf{RA} & 29  & 17   &\highlight{6}  &0.74       &0.61  &0.67 & +0.09  \\
\bottomrule
\end{tabular}
\renewcommand{\tabcolsep}{\oldtabcolsepaaaa}
\end{table}

\begin{table}[t]
\centering
\caption{SZZ variants performance (Part B). Note: inputs only run on SZZ.}
\label{tab:all_pfaszz}
\rowcolors{3}{white}{gray!10}
\newcommand{\oldtabcolsepaaaa}{\tabcolsep}
\renewcommand{\tabcolsep}{0.7em}
\begin{tabular}{l@{\hskip0.5cm}rrrrrr}
\toprule
& \multicolumn{6}{c}{\textbf{Part B (All Languages)}} \\
& \textbf{n} & \textbf{TP} & \textbf{FP} & \textbf{Recall} & \textbf{Precision} & $\boldsymbol{F_1}$ \\
\cmidrule(l{-0.5em}r{0em}){2-7}
\textbf{B} & 397 & 250 & 713 & 0.63 & 0.26 & 0.37   \\
\textbf{AG} & 397 & 185 & 300 & 0.46 & 0.38 & 0.42 \\
\textbf{L} & 397 & 107 & 165 & 0.27 & 0.39 & 0.32 \\
\textbf{R} & 397 & 161 & 106 & 0.40 & 0.60 & 0.48 \\
\textbf{MA} & 397 & 188 & 370 & 0.47 & 0.33 & 0.39  \\
\rowcolor{white}\\
\rowcolor{white}
 & \multicolumn{6}{c}{\textbf{Part B (Java)}} \\
\cmidrule(l{-0.5em}r{0em}){2-7}
\textbf{RA} & 31 & 8 & 23 & 0.26 & 0.26 & 0.26 \\
\bottomrule
\end{tabular}
\renewcommand{\tabcolsep}{\oldtabcolsepaaaa}
\end{table}

\begin{figure*}[ht]
  \centering
  \includegraphics[width=\linewidth]{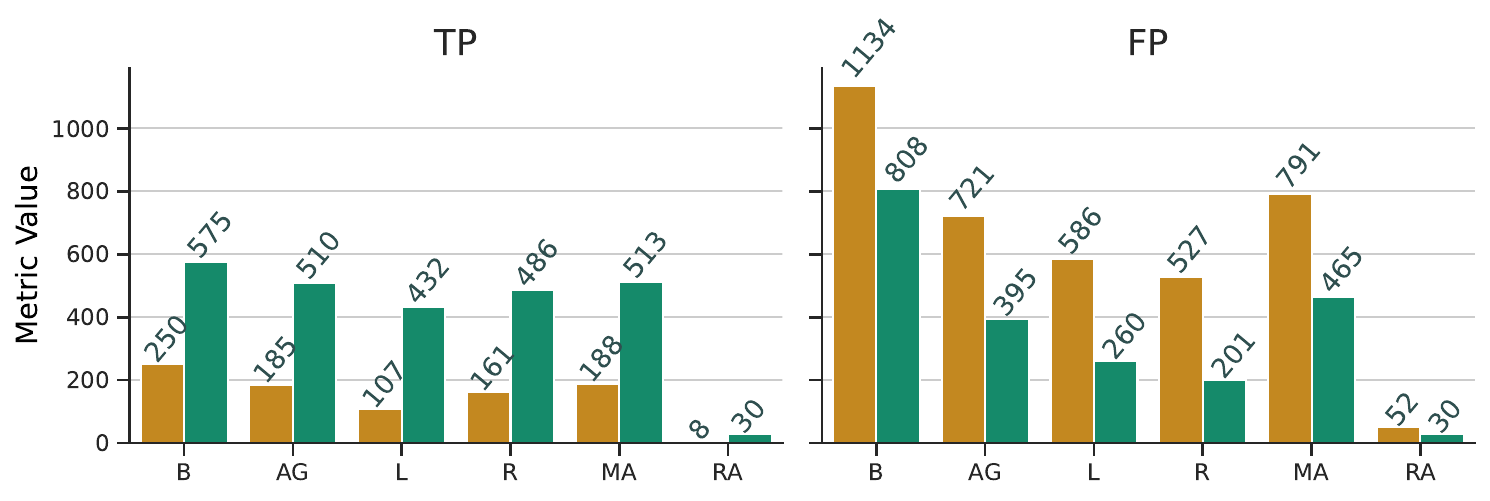}
  \includegraphics[width=\linewidth]{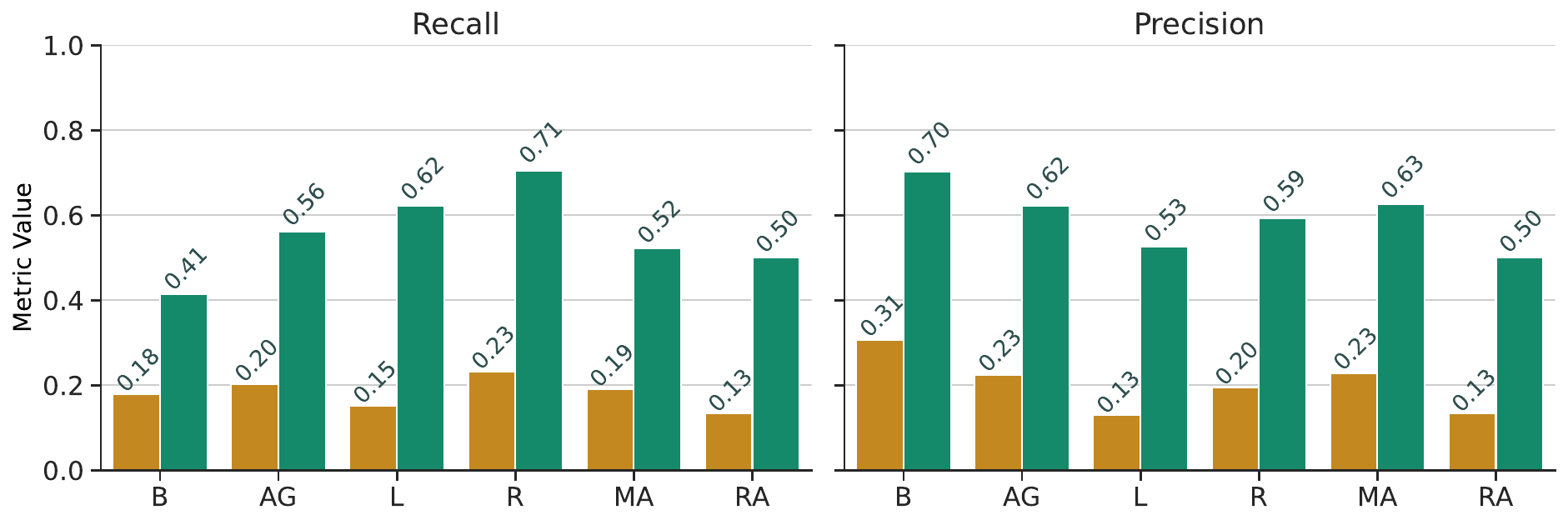}
  \includegraphics[width=0.8\linewidth]{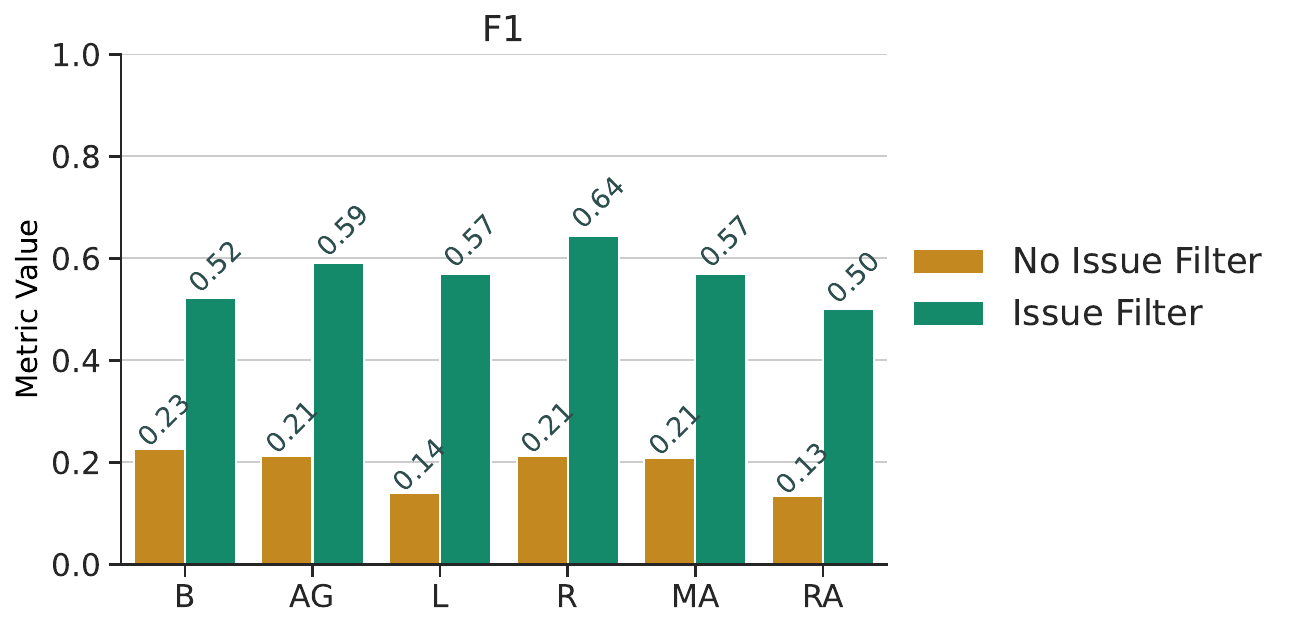}
  \caption{\pfaszz with \issuefilter (green bars) and without \issuefilter (orange bars). \issuefilter provides \pfaszz with context to set up a time window with respect to the issue report date and look for related work items.}
  \label{fig:wiaszz_2filters}
\end{figure*}

\subsubsection{Part B: FCs Without Work Items}
\label{subsec:partB}

Note that for Part B, the performance of \pfaszz is identical to those of the SZZ variants since \pfaszz defaults to calling the SZZ variants due to the lack of work items. Given this fact, it is clear the gains observed in \Cref{tab:all_pfaszzdeltas} come entirely from Part A. If future researchers wish to further improve SZZ, they could focus on Part B, as it currently performs poorly on average.

Now that we know how our approach compares to the SZZ variants, in the next sections we investigate the impact of some of the filters on these approaches.

\findings{
WIA-SZZ detects more BICs and fewer false positives, leading to a 5\% average increase in F1 scores across six SZZ variants, with the highest recall boost of 11\% when combined with L-SZZ. When work item groupings can be detected from experimental objects, WIA-SZZ outperforms SZZ variants with an F1 score increase from +3\% up to +14\%. Also, WIA-SZZ can identify BICs in scenarios where existing SZZ variants fail, such as when the FC includes only code insertions.
}

\subsection{\ref*{rq:filter}: Does applying an issue-date filter improve the accuracy of SZZ and/or \pfaszz?}

As noted by \cite{rosa21}, the \issuefilter reduces false positives in SZZ without affecting recall. However, its impact on our approach, \pfaszz, remains unclear. In this section, we explore the \issuefilter's effect on \pfaszz.

We conducted experiments with and without the \issuefilter enabled in \pfaszz. Without the filter, \pfaszz selects the first work item found before the FC, still returning only one BIC candidate per object. Results in \Cref{fig:wiaszz_2filters} indicate a significant negative impact on \pfaszz performance when the \issuefilter is disabled.

The absence of the filter leads \pfaszz to likely select supplementary fixes rather than the bug-inducing commit itself, resulting in mostly false positives and a notably lower recall rate, peaking at only 23\% without the filter. Precision also declines, reaching a peak of 31\%, with the highest $F_1$ at 23\%. These metrics are substantially lower than those with the filter enabled, where both recall and precision reach 70\%.

\findings{
The \issuefilter enhances \pfaszz performance, akin to findings by \cite{rosa21}. This underscores the importance of issue reports for \pfaszz effectiveness. In their absence, simulating issue report dates through sensitivity analysis with varying threshold values between FC and BIC dates could be beneficial, as demonstrated in this study.
}

\begin{figure*}[ht]
  \centering
  \includegraphics[width=\linewidth]{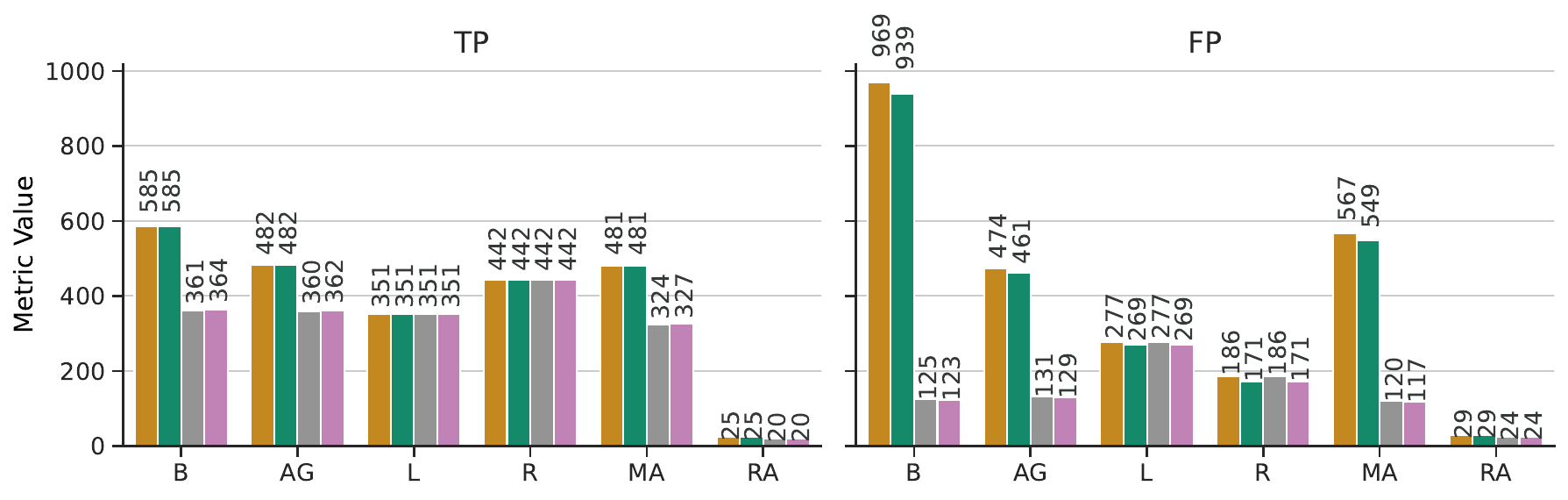}
  \includegraphics[width=\linewidth]{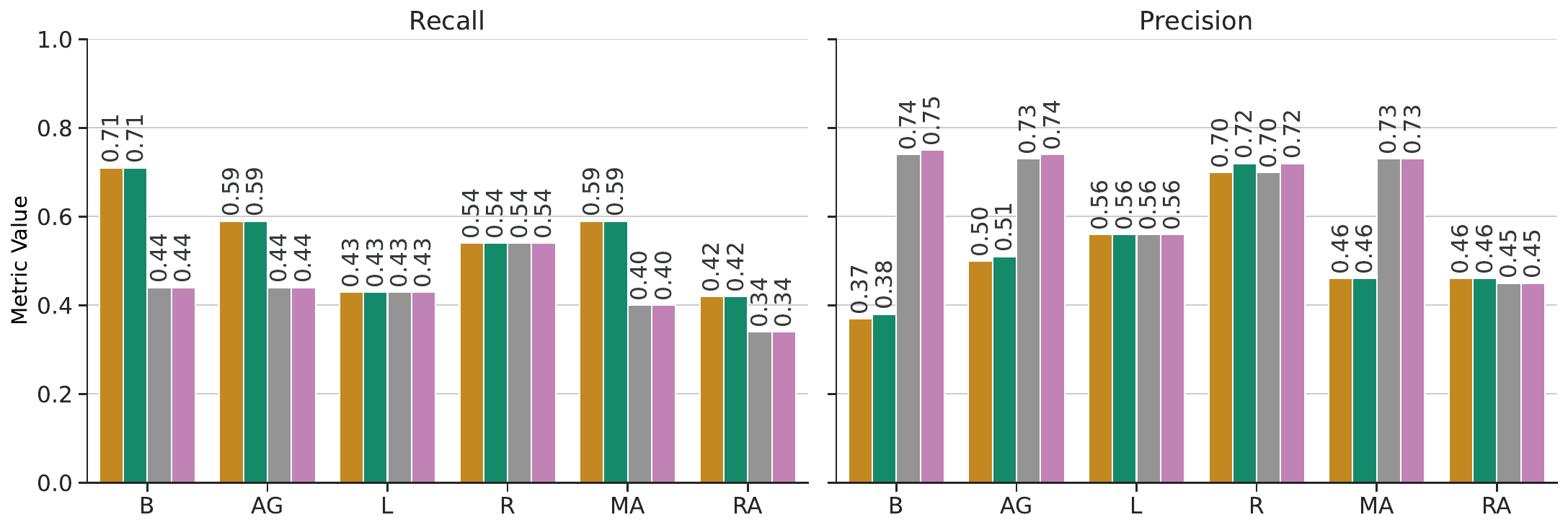}
  \includegraphics[width=0.9\linewidth]{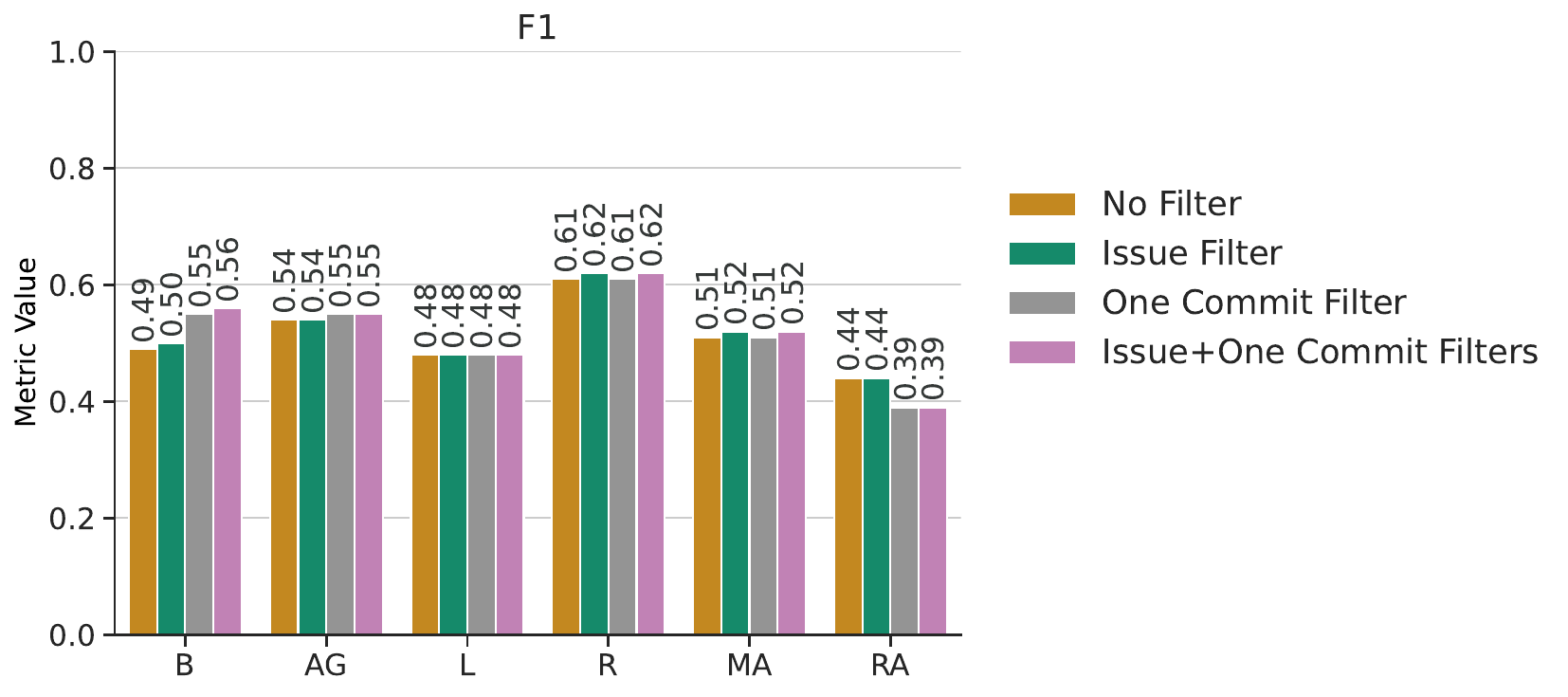}
  \caption{SZZ evaluation dataset metrics with No Filter (dark yellow bars), \issuefilter (green bars), \ocfilter (grey bars), and Issue + One Commit Filters (orchid bars). Note that other than the B variant, the filters do not significantly improve $F_1$ scores.}
  \label{fig:szz_4filters}
\end{figure*}

\subsection{\ref*{rq:overfit}: Does always selecting the candidate nearest the issue date cause overfitting?}
\label{subsec:overfit}

As mentioned in \Cref{sec:threats}, the use of the \ocfilter can result in overfitting because it picks only one candidate just prior to the issue report date.  In the worst case, if the issue report is always immediately after the bug-inducing commit, such an approach would almost certainly yield near-perfect results (e.g., if we used simulated issue dates of 60 seconds after the BICs).  This is because the choice will always yield one BIC candidate, minimizing the false positive rate.  If the choice is always correct, it would also maximize the true positive rate.  In this section, we investigate if the use of this filter falls into overfitting.

To investigate the impacts of the \ocfilter, we compared the performance of the SZZ variants with and without the \ocfilter applied.  We also investigated the impacts of the \ocfilter when used with and without the \issuefilter enabled. We show the results of this analysis in \Cref{fig:szz_4filters}.

In this table, we show the results for SZZ variants with no filters applied, the \issuefilter applied, the \ocfilter applied, and with both filters applied.  Similar to the prior results~\citep{rosa21}, we see that the use of the \issuefilter keeps the recall exactly the same for all SZZ variants, while also slightly improving the precision (by 0-2\%).

Regarding the \ocfilter, it notably boosts overall precision as expected by reducing multiple candidate BICs to at most one, thus enhancing accuracy. However, its impact on recall varies: in four cases, recall decreases due to incorrect candidate selection, while in two cases, it remains unchanged. Overall, accuracy is similar to that with the issue filter.

In the case of the F1 performance of B-SZZ in \Cref{fig:szz_4filters}, it is noteworthy that the performance with the \ocfilter enabled outperforms \pfaszz.  This was because \pfaszz used the original B-SZZ, which does not enable this filter.  Generally, any improvement applied to SZZ variant(s) could also improve the performance of \pfaszz when falling back to process data in Part B \Cref{subsec:partB}.
\begin{samepage}
\findings{
We conclude that applying the \ocfilter does not overfit the results, and it is safe to use such a filter in our approach.  We can also conclude that, at least for SZZ, it does not seem to benefit their approaches unless the SZZ variant already has a high false positive rate.
}
\end{samepage}

\section{Threats to Validity}
\label{sec:threats}

Concerning \textbf{internal validity}, a primary issue arises from the differing approaches of \pfaszz and SZZ variants in reporting BIC candidates. While \pfaszz reports one candidate per object, SZZ variants can identify multiple candidates. This raises the question of whether integrating the \ocfilter could improve SZZ variant performance by reducing false positives. To address this, we investigated the impact of each filter on both approaches, evaluating SZZ variants under different conditions and assessing the effect of \issuefilter on \pfaszz performance.

Concerning the 397 FCs that do not have work items, we analyzed common trends among them. First, 149 FCs lacked modified methods at the fix commit, except for tests, preventing \pfaszz from generating tracking matrices; thus, there were no work items. Second, 106 FCs had no overlapping methods between the FC and BIC, making it infeasible for our approach to find the BIC. Lastly, we identified 244 FCs with a BIC outside the time window we used to look for work items.

Regarding \textbf{external validity}, a notable issue involves the inherent subjectivity in our manual evaluation process due to biases inherent in human judgments. To mitigate this, we introduced a third rater with extensive expertise. This rater validated judgments, particularly concerning the most intricate part of the evaluation sample—namely, the work items from FCs that do not include the oracle BIC.

Another concern is the representativeness of the dataset in \cite{rosa21}. It includes various project sizes and file counts across eight popular languages, making it representative of real-world software engineering projects.

A third threat involves the generalizability of our filtering factor. Results in \Cref{tab:all_pfaszz} show minimal variations in performance metrics across different factors, allowing users to tailor filtering based on priorities. For example, users of our system can choose to prioritize maximizing the number of TP (by selecting a lower factor), minimizing the number of FP (by opting for a higher factor), or achieving a modest change in TP but fewer FP (with an intermediate factor).

\section{Conclusion}
\label{sec:conclusion}

We present \pfaszz, a method that leverages work items to enhance the effectiveness of SZZ variants. Our experiments, conducted on publicly available repositories commonly used in SZZ-related studies, demonstrate that our approach can increase $F_1$ scores by 2\% to 9\% (and 3\% to 14\% on cases where a work item can be identified), providing solid evidence of bug-fix associations. Additionally, we evaluated two filters to gauge their impact on overall performance. The \ocfilter notably reduces false positives for SZZ but also misidentifies true positives, diminishing SZZ's overall performance. However, \pfaszz achieves optimal performance when both the \ocfilter and the \issuefilter are utilized.

For future work, we plan to identify repository instances unsuitable for our work item-based heuristic and uncover patterns to enhance SZZ's effectiveness in identifying BICs. Moreover, we plan to employ the work item-based heuristic to tackle other research challenges, such as refining commit histories for research datasets and characterizing repairs/defects.

\section*{Data Availability}
The prototype implementation of \pfaszz and the dataset used for evaluation are available on Zenodo~\citep{replication-package}.

\section*{Conflict of Interest}

The authors declare that they have no conflict of interest.

\begin{acknowledgements}
  This work was supported by the U.S. National Science Foundation under the grant CNS-2346327.
\end{acknowledgements}

\bibliographystyle{spbasic}
\bibliography{refs.bib}
\end{document}